\documentclass{aa}
\usepackage{psfig} 
\begin{document}
\thesaurus{}
\title{Group analysis in the SSRS2 catalog }

\author{C.~Adami \inst{1}, A.~Mazure \inst{1}}
\institute{LAM, Traverse du Siphon, F-13012 Marseille, France}
\offprints{C.~Adami} 
\date{30 Juillet 2000, Accepted     } 
\maketitle 
\markboth{Group analysis in the SSRS2 catalog}{} 

\begin{abstract} 

We present an automated method to detect populations of groups in galaxy
redshift catalogs. This method uses both analysis of the redshift distribution
along lines of sight in
fixed cells to detect elementary structures and a friend-of-friend algorithm
to merge these elementary structures into physical structures. We apply this
method to the SSRS2 galaxy redshift catalog. The groups detected with our
method are similar to group catalogs detected with pure friend-of-friend
algorithms. They have similar mass distribution, similar abundance versus
redshift, similar 2-point correlation function (modeled by a power law: $
(r/r_0)^\gamma $ with $r_0=7h^{-1}Mpc$ and $\gamma =-1.79$) and the same
redshift completeness limit, close to 5000 km/s. If instead of SSRS2, we use
catalogs of the new generation (deep redshift surveys obtained with 10
meters class telescopes), it would lead to a completeness limit of z$\sim$0.7.

We model the luminosity function for nearby galaxy groups by a Schechter
function with parameters M$_{SSRS2}^{*}$=(-19.99$\pm $0.36)+5logh and $%
\alpha $=-1.46$\pm $0.17 to compute the mass to light ratio. The median
value of the mass to light ratio is 360 hM$_{\odot }$/L$_{\odot }$ (in
the SSRS2 band, close to a B band magnitude) and we deduce a relation
between mass to light ratio and velocity dispersion $\sigma $ ($%
M/L=(3.79\pm 0.64)\sigma -(294\pm 570)$). The more massive the group, the
higher the mass to light ratio, and therefore, the larger the amount of dark
matter inside the group. Another explanation is a significant stripping of
the gas of the galaxies in massive groups as opposed to low mass groups.
This extends to groups of galaxies the mild tendency already detected for
rich clusters of galaxies. Finally, we detect a barely significant
fundamental plane for these groups ($L\propto R_{Vir}^{2.26\pm 1.39}\times
\sigma ^{2.93\pm 1.64}$ for groups with more than 8 galaxies) but much less
narrow than for clusters of galaxies.

\end{abstract}


\section{Introduction}

The abundance evolution of galaxy structures is a major prediction of
cosmological models (e.g. Oukbir $\&$ Blanchard 1992, 1997, Romer et al.
2001). The more distant the structure, the strongest the constraint. Until
now, however, we are limited to the study of the most massive distant
structures. This is because these structures are the easiest to detect at
high redshift using for example X-ray selected samples (e.g. Borgani et al. 
1999, Burke et al. 1997,  Ebeling et al. 2000, Gioia et al. 1990, Romer et 
al. 2000, Vikhlinin et al. 1998). These distant massive structures are the
progenitors of the most massive z$\sim $0 clusters. The hierarchical models
predict, however, that nearby intermediate mass clusters have formed from
low mass high redshift systems as groups of galaxies. The new generation of
deep galaxy surveys (VLT and Keck surveys) will be able to detect such high
redshift low mass structures. This will add an element to the large scale
structure formation models.

We present in this paper a new efficient method to detect and study such
structures in very large galaxy redshift samples (Section 2). We applied
this method to the SSRS2 (da Costa et al. 1998) galaxy redshift catalog in
order to quantify our detection efficiency at low redshift. We have created
a catalog of groups from this survey and we have studied the properties of
these structures (Section 3), comparing our results to literature studies
(Carlberg et al. 2001, Girardi et al. 2000, Merchan et al. 2000). We
conclude in Section 4.

\section{The method}

The structure detection we used is an hybrid method derived from the method
used for the ENACS\ sample of galaxy clusters (e.g. Katgert et al. 1996) and
from classical friend-of-friend detection methods (e.g. Huchra $\&$ Geller
1982). The friend-of-friend algorithm is applied to a preprocessed catalog
of elementary galaxy associations and not to individual galaxies. In this way we 
reduce the computation time by a large factor (see Section 3.3). The procedure
consists of four steps:

\begin{itemize}
\item  We apply a running window (with an overlap of the half of the window
size) to the original catalog of galaxies to divide the total area covered
by the survey in several elementary lines of sight. The angular size $%
\varepsilon $ of the running window is chosen as a function of the kind of
structures we are trying to detect ($\varepsilon $ has to be close to the
typical size of the structures) and as a function of the redshift of these
structures.

\item  We search for compact galaxy redshift associations in each of these
elementary beams by using a technique similar to Katgert et al. (1996).
This technique is basically detecting gaps (fixed redshift separations
between successive galaxies) greater than a given value (noted ``gap analysis'' in
the following). This value is given by the typical maximal
velocity dispersion of the structures we try to detect. For groups, we
searched galaxies separated by more than 600 km/s in each elementary beam.
If two galaxies are separated by more than this value, these galaxies belong
to different structures. An elementary structure has more than 2 galaxies.

\item  We apply a friend-of-friend algorithm to merge the elementary
structures into larger real structures. Two elementary structures were
assumed to be connected if they were closer than $\varepsilon $. This value
is an approximation of the maximal diameter of a group in real space. This
allows to generate a preliminary catalog of interesting regions with
potential structures inside.

\item  Finally, we re-apply the gap method for each of these regions in
order to detect real structures in the catalog.
\end{itemize}

\section{Application to the SSRS2 catalog: group detection and analysis}

The SSRS2 catalog is a sample of 5369 galaxies, covering a region of 1.70 sr
of the southern sky ($b\leq -40\deg $ and $\delta \leq -2.5\deg $ for the
southern galactic cap and $b\geq 35\deg $ and $\delta \leq 0\deg $ for the
northern galactic cap ). This catalog is 99\% complete down to m$%
_{SSRS2}=15.5$. The m$_{SSRS2}$ magnitude is close to a B magnitude (see da
Costa et al. 1998). The precision of the individual galaxy redshift
measurements is $\sim $40 km/s. Detailed information on this catalog can be
found in da Costa et al. (1998). We limited our analysis to the
2000$<$cz$<$20000 km/s range. To detect groups, we used $\varepsilon
=1.55h^{-1}Mpc$. This value is an approximation of the maximal diameter of a
group and gives the most similar results compared to the SSRS2 group
analysis by Merchan et al. (2000). We kept only the structures with more
than 4 galaxies and less than 40 galaxies (to be consistent with Merchan et
al. 2000). Structures with less than or with 3 galaxies are possibly close
superposition effects and not real dynamical structures (Ramella et al.
1997). Structures with more than 40 galaxies are likely to be rich clusters
or diffuse elongated filaments of galaxies.

\subsection{Richness and physical extension}

The detected structures are described in Table 1 and 2. The mean number of
galaxies per structure is $6\pm 5$ and 90\% of the structures have less than
11 galaxies. The largest group has 32 galaxies. The richest groups are the
closer ones, in agreement with the detection of a uniform richness class of
groups. This is because, due to apparent limiting magnitude, the closest
groups have more galaxies brighter than the SSRS2 catalog magnitude limit.

The groups are more or less circular: they have the same angular extension
in $\alpha $ (right ascension) and $\delta $ (declination). We have $\Delta
\alpha -\Delta \delta =(-0.09\pm 0.26)$arcmin (computed with 80 groups).
This means that we are not detecting systematically elongated structures
such as poor filaments. The mean extension of the detected groups (computed
as the dispersion of the coordinates of the galaxies of a given group) is $%
(0.35\pm 0.18)$ $h^{-1}Mpc$. We have no variation
of the group size as a function of the redshift, but the richest groups are
also the larger ones. The groups with more than 10 galaxies have a size in
the range [0.45;0.8] $h^{-1}Mpc$. The groups with less than 10 galaxies have
a size in the range [0.05;0.65] $h^{-1}Mpc$. Finally, the mean virial radius
R$_{Vir}$ of the groups is $(0.37\pm 0.21)h^{-1}Mpc$ (in the range
[0.06;0.87] $h^{-1}Mpc$, see Table 1 and 2).

\begin{figure} 
\vbox 
{\psfig{file=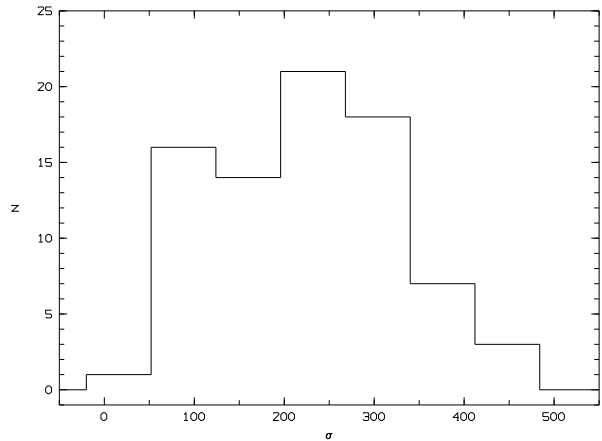,width=9.0cm,angle=270}} 
\caption[]{Histogram of the group velocity dispersions (in km/s).} 
\label{} 
\end{figure}

\subsection{Velocity dispersions and masses}

We computed velocity dispersions using a robust estimator (Beers et al.
1990). This gives a more realistic estimate of the velocity dispersion for
groups with more than 10 galaxies, but still a low number, compared to
classical estimators. For groups with less than 10 galaxies, there is no
significant difference between this robust estimator and the classical
estimator of the velocity dispersion (Lax 1985).

The mean velocity dispersion of the groups is $(224\pm 103)$km/s. None of
them has a velocity dispersion greater than 500 km/s (see Fig 1). These
values are consistent with several other computations for groups (e.g.
Carlberg et al. 2001 with a median velocity dispersion of 200 km/s). This
means that we detected no rich clusters. The groups with the highest
velocity dispersions are also the richest (see Fig 2) but we have no
significant relation between the group mean redshift and velocity
dispersion. This is, one more time, in favor of a uniform detection of
groups as a function of redshift.

In order to check how many spurious groups we are detecting (groups with a
large velocity dispersion, but a low number of galaxies: chance alignments),
we plotted the velocity dispersion as a function of the number of galaxies
inside the groups (Fig 2). Only 10\% of the groups have less than 5 galaxies
and a velocity dispersion greater than 300 km/s. This can be assumed to be
the contamination rate of our catalog.

Combining the virial radius and the velocity dispersion, we computed the
mass using the standard virial mass estimator (M $\propto $ virial radius 
$\times $ (velocity dispersion)$^2$) (e.g. Ramella et al. 1997). About 85\%
of the sample has a mass in the range [$5.10^{12}$;$4.10^{14}$] solar
masses and 97.5\% has a mass in the range [$10^{12}$;$4.10^{14}$] solar
masses. None of the groups has a mass greater than $4.10^{14}$ solar masses.
The mean mass is ($5.6\pm 6.5$)$10^{13}$ solar masses. All the masses
computed by Merchan et al. (2000) are in the range [$5.10^{12}$;$4.10^{14}$%
] solar masses. This means that we have slightly lower masses in the sample
(for 15\% of the sample). This is probably due to the fact that we are using
robust estimators to compute velocity dispersions. Such estimators are less
biased toward high values for small samples of galaxies. The velocity
dispersions and masses are given in Table 1 and 2.

\begin{figure} 
\vbox 
{\psfig{file=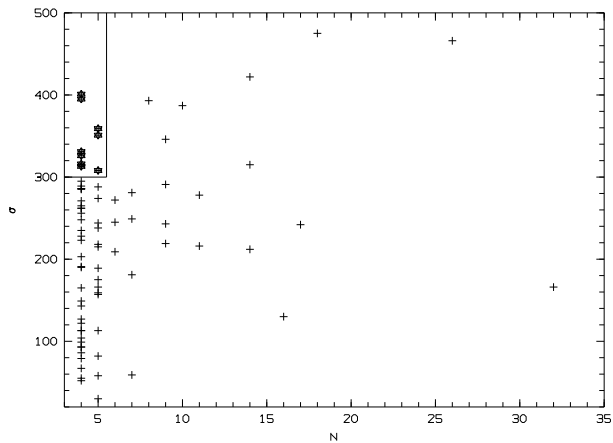,width=9.0cm,angle=270}} 
\caption[]{Number of galaxies inside each group as a function of the velocity dispersion of the group. The stars are the groups with too low a richness compared to the velocity dispersion.} 
\label{} 
\end{figure}

In order to estimate the mass uncertainty, we used galaxy groups with more
than 10 galaxies. For these groups, a reliable error can be computed for
velocity dispersion using 100 bootstrap resamplings. This is obviously a
lower limit of the velocity dispersion uncertainty as this uncertainty is
probably larger for smaller groups. The 1-$\sigma $ uncertainty is: $\Delta
\sigma /\sigma \sim 20\%$. The mass uncertainty can be written as:

\[
\Delta Mass/Mass=\Delta R_{Vir}/R_{vir}+2\Delta \sigma /\sigma 
\]

Assuming that the velocity dispersion uncertainty is the dominant factor, we
have the 1-$\sigma $ error on the mass: $\Delta Mass/Mass\geq 40\%$. This
large error raises the question of the validity of virial mass estimates for
groups of galaxies as these structures are probably not all in equilibrium
(e.g. Dos Santos $\&$ Mamon 1999).

\subsection{Sample completeness and 2-point correlation function}

We compared our catalog with the one of Merchan et al. (2000) that used the
same galaxy sample. Fig 3 shows the group redshift distribution of the two
catalogs. We are detecting similar numbers of groups up to cz$\sim $8000
km/s. The mean difference per bin is 5\% (with redshift bins of 1250 km/s
width). Merchan et al. are detecting more groups at higher redshift.
However, their catalog begins to be incomplete above 5000 km/s. This means
that we have the same detection completeness level, in agreement with Press
$\&$ Schechter (1974) predictions (Merchan et al. 2000). Finally, we estimated
the 2-point correlation function for the sample. Using a power law
approximation of the form $(r/r_0)^\gamma $, we have $r_0=7.0h^{-1}Mpc$ and $%
\gamma =-1.79$. This is consistent at the 1$\sigma $ level with the values of
Merchan et al. (2000): $r_0=(8.4\pm 1.8)h^{-1}Mpc$ and $\gamma =-2.0\pm 0.7$%
, Girardi et al. (2000): $r_0=(8.\pm 1.)h^{-1}Mpc$ and $\gamma =-1.9\pm 0.7$
or Carlberg et al. (2001): $r_0=(6.8\pm 0.3)h^{-1}Mpc$ and $\gamma =-1.8$
(fixed). We detect a positive signal up to $\sim $90 $h^{-1}Mpc$ (similar to
the values of Merchan et al. 2000). These estimates are all in good
agreement and give confidence in the detection method. The only significant
difference is a lower group detection rate at the faint end of the galaxy
catalog. However, this is not affecting the completeness level of our group
sample as we detect the same number of groups up to cz$\sim $8000 km/s
(assuming that the Merchan et al. completeness limit is correct).

\begin{figure} 
\vbox 
{\psfig{file=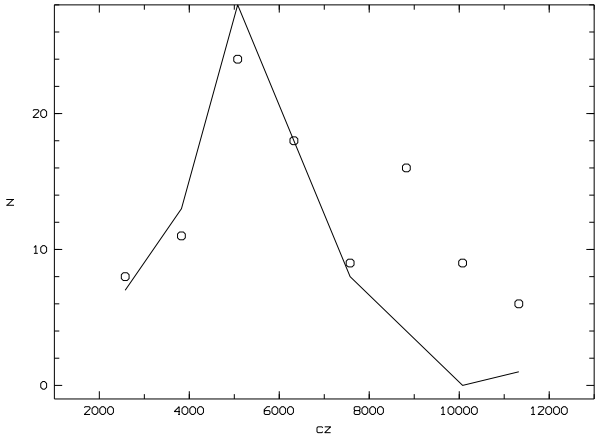,width=9.0cm,angle=270}} 
\caption[]{Number of detected groups as a function of redshift by Merchan et al. (circles) and with our method (solid line).} 
\label{} 
\end{figure}

The redshift completeness limit close to 5000 km/s is directly related to
the magnitude limit of the catalog: m$_{SSRS2}\sim B\sim 15.5$. It gives an
absolute magnitude of -18+5logh. This is the magnitude limit for the
galaxies lying inside groups before the completeness limit. We can
extrapolate this redshift completeness limit assuming the typical depth of
the new generation of redshift surveys using 10 meter class telescopes
(Keck and VLT). For example, the deep part of the Virmos survey should
provide a redshift survey 80\% complete down to B$\sim $26, and the shallow
part of this survey a 50\% complete redshift catalog down to B$\sim $24.
Assuming a k-correction proportional to the redshift, it should provide a
catalog of groups complete up to z$\sim $1.2 for the deep catalog and z$\sim 
$0.7 for the shallow catalog. This is enough to sample nearly all classes of
galaxy structures up to z$\sim $0.7. This will allow to put strong
constraints on cosmological models (e.g. Romer et al. 2001) as well as
sampling the internal structure evolution of groups (and more massive
clusters) up to these redshifts.

The time required to treat these future samples with our method
is very short. We used the simulations of Steve Hatton (private communication)
in order to estimate these times. These simulations cover 1 deg$^2$ and
reproduce an $\Omega_m$=0.3 and $\Omega_\Lambda$=0.7 Universe, including
several structures as clusters or groups. We used a single 6' running window 
for simplicity and we selected randomly different sub-samples of galaxies in 
this catalog. The computation times are given in Fig 4. We see that a catalog
of 150000 galaxies, comparable to the spectroscopic catalogs which will be 
produced by the VLT and Keck surveys, is completely analyzed in about 2 
minutes (using an ES40 Compaq, processor EB67 at 600 Mhz).

\subsection{Luminosity function}

We computed the luminosity function of the group galaxies of the detected
sample. The galaxy catalog is complete down to m$_{SSRS2}=15.5$ with groups
up to 8000 km/s. Limiting ourselves to groups with velocity lower than 6000
km/s (this velocity limit is close to the completeness limit of the group
sample), we have a galaxy catalog complete down to M$_{SSRS2}$=-18.4+5logh.
Using the estimate of Zabludoff $\&$ Mulchaey (2000), this corresponds to
about M$^{*}$+1.5. We fitted a Schechter function on the magnitude
distribution down to M$_{SSRS2}$=-18.4+5logh and for groups with cz$\leq $%
6000 km/s using a maximum likelihood technique (e.g. Lobo et al. 1997).

The absolute magnitudes were k-corrected (but this is a minor correction due
to the low redshift of the sample: see Table 1 and 2) using the k-correction 
-3.5$%
\times $z (Rauzy et al. 1998). We also corrected for galactic extinction
using the work by Schlegel et al. (1998). The correction we applied for the
magnitudes was -4.325$\times $E(B-V). The mean correction was 16\% of the
luminosity (but up to 60\% for the worse cases: see Table 1 and 2).

The best fit parameters of a Schechter function are M$^{*}$=(-19.99$\pm $%
0.36)+5logh and $\alpha $=-1.46$\pm $0.17 (estimates with 1-$\sigma $
error). This is consistent at the 1-$\sigma $ level with the estimates of
Zabludoff $\&$ Mulchaey (2000): M$_B^{*}$=(-20.1$\pm $0.4)+5logh (assuming m$_B
$-m$_R\sim $1.5) and $\alpha $=-1.3$\pm $0.1 for groups in
2800$<$cz$<$7700 km/s. These values are also similar at the 1-$\sigma $ level
with the estimates of Rauzy et al. (1998) for rich clusters of galaxies.

In order to check the robustness of our estimates, we split the sample in
two parts: the northern and the southern galactic cap. The southern cap
gives M$^{*}$=(-20.26$\pm $0.69)+5logh and $\alpha $=-1.61$\pm $0.47 and the
northern cap gives M$^{*}$=(-19.79$\pm $0.45)+5logh and $\alpha $=-1.32$\pm $%
0.42. Despite the larger uncertainty, these values are still consistent,
with, however, a mild tendency to have more faint galaxies in the southern
galactic cap.

\subsection{Total luminosity and mass to light ratio}

We computed the total luminosity of each group summing up all the individual
magnitudes (see Table 1 and 2). We used M$_{SSRS}^{\odot }$=5.53 
(approximation of
M$_B^{\odot }$). We corrected these values for incompleteness due to galaxy
catalog magnitude limit by using the ratio between the luminosity function
integrated from M$_{SSRS}^{}$=-23+5logh. to -10+5logh and integrated between
the faintest magnitude of each group and M$_{SSRS}^{}$=-10+5logh. This is
because we assumed that the faintest galaxies were not fainter than
-10+5logh. This limit has, however, a moderate influence on this correction
factor. We used the luminosity function computed in the previous section
with M$^{*}$=(-19.99$\pm $0.36)+5logh and $\alpha $=-1.46$\pm $0.17. The
mean correction was 135\% of the raw group luminosity. The total
luminosities and the individual completeness corrections are given in Table
1 and 2.

Combining estimates of the mass and of the luminosity, we computed the mass
to light ratio (see Table 1 and 2). The mean value is 550 hM$_{\odot }$/L$%
_{\odot }$ and the median is 250 hM$_{\odot }$/L$_{\odot }$. We compared
these estimates with those of Carlberg et al. (2001). In the R band, they
computed median values in the range [150;250] hM$_{\odot }$/L$_{\odot }$.
Assuming m$_{R\odot }$-m$_{SSRS\odot }$=-1.17 (Allen 1973) and m$_R$-m$%
_{SSRS}$=-1.5 for galaxies in nearby structures (e.g. Katgert et al. 1998),
it corresponds to median values in the range [204;340] hM$_{\odot }$/L$%
_{\odot }$ in the SSRS2 magnitude passband (multiplication factor of 10$%
^{0.4(1.5-1.17)}$). This is in good agreement with our estimates. We found a
relation between the mass to light ratio and the velocity dispersion $\sigma 
$ for our sample (see Fig. 5 and Table 3).

\begin{figure} 
\vbox 
{\psfig{file=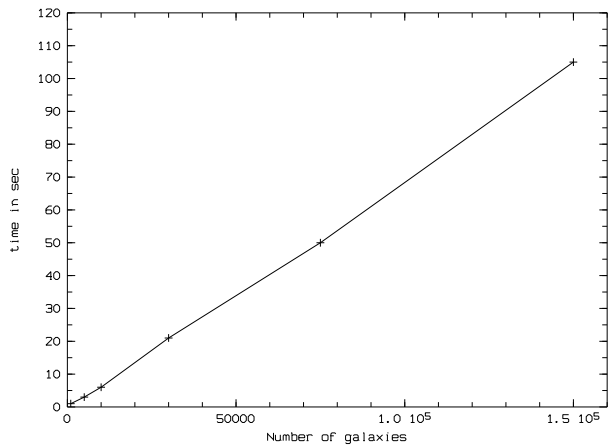,width=9.0cm,angle=270}} 
\caption[]{Computation time (in seconds) for a simulated catalog of galaxies 
including groups and clusters, as a function of the number of galaxies inside 
the catalog.} 
\label{} 
\end{figure}

These relations are qualitatively similar to the relation we found between
rich cluster velocity dispersion and cluster mass to light ratio (Adami et
al. 1998): $M/L=(0.58\pm 0.17)\sigma -(12\pm 236)$. The slope is,
however, significantly steeper for groups than for clusters. The slope of
these relations is also similar to that of Carlberg et al. (2001) which is
close to 3.5.\ The more massive the group (or the cluster), the higher the
mass to light ratio, and therefore, the larger the amount of dark matter
inside the group (or cluster). Another explanation would be a very efficient
stripping of the gas of the galaxies in rich groups (as opposed to very low
mass groups). This would be due to a denser intra group medium or a more
efficient tidal stripping due to the higher number of galaxies (e.g. Dos
Santos $\&$ Mamon 1999) in rich groups. It would lower the star formation rate
in galaxies, inducing a higher mass to light ratio for the richest groups
(see also Carlberg et al. 2001). The dispersion of the relation is also
smaller for richer groups. This is probably due to a better computation of
the velocity dispersion due to the larger number of available galaxies. 

\begin{figure} 
\vbox 
{\psfig{file=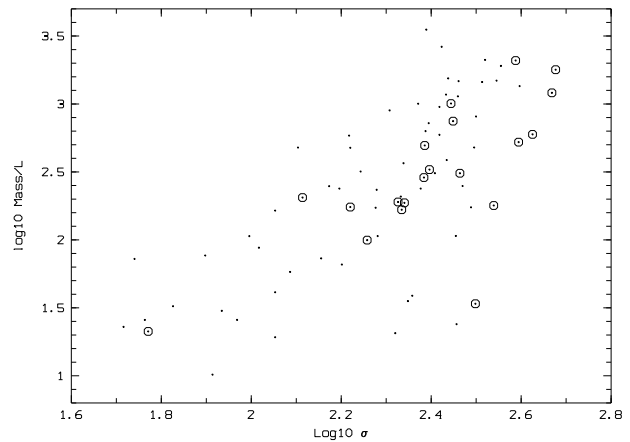,width=9.0cm,angle=270}} 
\caption[]{Mass to light ratio as a function of the velocity dispersion for 
groups of galaxies detected by our method in the SSRS2 galaxy catalog. The 
circles are the groups with more than 8 galaxies.} 
\label{} 
\end{figure}

\subsection{A fundamental plane for groups of galaxies?}

Schaeffer et al. (1993) or Adami et al. (1998) have shown, using optical
data, that clusters of galaxies populate a well defined area in the [total
luminosity: L, characteristic radius of a beta profile: R$_c$, velocity
dispersion: $\sigma $] space. Adami et al. (1998) found the following
relation:

L$\propto$ R$_c^{1.19}$$\times$$ \sigma ^{0.91}$ with a dispersion of the
relation of 5$\%$

Using the virial radius instead of the characteristic radius of a beta
profile would give:

L$\propto$ R$_{Vir}^{1.51}$$\times$$ \sigma ^{1.07} $

This relation called the fundamental plane is also detected for elliptical
galaxies, but with different coefficients (see Adami et al. 1998 for a
comparison). This is interpreted as the different relaxation state of
elliptical galaxies and clusters of galaxies. We searched for the same kind
of relation in the groups we detected. Using all the groups, we found no
relation between L, R$_{Vir}^{}$ and $\sigma $. Using only
groups with more than 8 galaxies, we found:

L$\propto$ R$_{Vir}^{2.26\pm 1.39}$$\times$$ \sigma ^{2.93\pm 1.64}$ with a
dispersion of 58$\%$

The coefficients are consistent with those of clusters of galaxies, but the
uncertainties are very large, and the intrinsic dispersion of the relation
is 10 times larger than for clusters. We conclude that a mean relation
between L, R$_{Vir}$ and $\sigma $ probably exists for groups, but this
relation is much less well defined than for rich clusters. Moreover, poor
groups (less than 7 galaxies) probably have too large a velocity dispersion
uncertainty to allow any detection of this relation.

\begin{table*} 
\caption[]{Group sample: col1 (N): number of galaxies inside groups, col2 ($\alpha$): right ascension (in degrees), col3 ($\delta$): declination (degrees), col4 (z): mean redshift, col5 (Obs. Lum.): observed group luminosity (luminosity before corrections: in unit of solar luminosity), col6: E(B-V) from Schlegel et al. (1998) at the group coordinates, col7 (Ext.): extinction correction factor, col8 (Comp.): completeness correction factor, col9 (Tot. Lum.): group total luminosity (in unit of solar luminosity), col10 (Mass): virial mass (in unit of solar mass), col11 (M/L): mass to light ratio in the SSRS2 magnitude passband, col12 ($\sigma$): velocity dispersion in km/s, col13 (Rvir): virial radius in $h^{-1}Mpc$.}
\begin{flushleft} 
\begin{tabular}{crrccrccccrrr} 
\hline 
\noalign{\smallskip} 
N	&	$\alpha$ 	&	$\delta$ 	&	z	&
	Obs. Lum.	&	E(B-V)	&	Ext.	&	Comp.	&
	Tot. Lum.	&	Mass	&	M/L	&	
$\sigma$	&	Rvir	\\
\hline 
\noalign{\smallskip} 
5	&	3.2488	&	-24.3504	&	0.0255	&	5.33E+10	&	0.02220	&	1.09	&	3.20	&	1.86E+11	&	1.90E+12	&	10.2	&	82	&	0.13	\\
4	&	3.2868	&	-7.4360	&	0.0181	&	4.54E+10	&	0.03389	&	1.14	&	1.94	&	1.01E+11	&	4.91E+13	&	486.9	&	286	&	0.27	\\
4	&	5.5908	&	-4.3517	&	0.0138	&	1.13E+10	&	0.03145	&	1.13	&	1.60	&	2.04E+11	&	2.13E+14	&	1043.7	&	401	&	0.6	\\
5	&	6.8999	&	-9.0373	&	0.0192	&	2.99E+10	&	0.03553	&	1.15	&	2.19	&	7.56E+10	&	1.44E+14	&	1905.7	&	359	&	0.51	\\
4	&	7.0446	&	-37.3208	&	0.0245	&	3.91E+10	&	0.02375	&	1.10	&	3.13	&	1.34E+11	&	4.04E+12	&	30.1	&	86	&	0.25	\\
4	&	7.1985	&	-22.9825	&	0.0268	&	7.17E+10	&	0.01305	&	1.05	&	3.06	&	2.31E+11	&	4.44E+12	&	19.2	&	113	&	0.16	\\
5	&	7.3508	&	-10.9007	&	0.0122	&	5.79E+10	&	0.03334	&	1.14	&	2.15	&	1.42E+11	&	2.45E+13	&	172.4	&	189	&	0.31	\\
5	&	8.0602	&	-3.3412	&	0.0196	&	3.70E+10	&	0.03581	&	1.15	&	2.23	&	9.51E+10	&	1.98E+13	&	208.3	&	215	&	0.2	\\
4	&	8.4047	&	-28.5031	&	0.0234	&	3.47E+10	&	0.02017	&	1.08	&	2.62	&	9.86E+10	&	2.45E+13	&	248.4	&	149	&	0.5	\\
4	&	8.8372	&	-23.7145	&	0.0128	&	1.90E+10	&	0.01679	&	1.07	&	1.80	&	3.65E+10	&	3.89E+12	&	106.5	&	99	&	0.18	\\
7	&	10.6675	&	-9.3056	&	0.0197	&	4.22E+10	&	0.03604	&	1.15	&	1.89	&	9.20E+10	&	6.88E+13	&	747.4	&	281	&	0.4	\\
4	&	12.7327	&	-13.4234	&	0.0376	&	9.85E+10	&	0.02390	&	1.10	&	7.94	&	8.60E+11	&	2.06E+13	&	24.0	&	286	&	0.12	\\
4	&	14.2872	&	-9.1834	&	0.0150	&	2.88E+10	&	0.08095	&	1.38	&	2.07	&	8.22E+10	&	1.35E+13	&	164.2	&	113	&	0.48	\\
9	&	18.0011	&	-32.3378	&	0.0192	&	1.25E+11	&	0.02365	&	1.10	&	2.18	&	3.00E+11	&	5.37E+13	&	179.0	&	346	&	0.2	\\
4	&	18.0096	&	-6.6967	&	0.0204	&	2.76E+10	&	0.11385	&	1.57	&	2.37	&	1.03E+11	&	1.10E+13	&	107.0	&	285	&	0.06	\\
14	&	20.7478	&	-35.3655	&	0.0193	&	1.23E+11	&	0.01880	&	1.08	&	2.17	&	2.88E+11	&	5.47E+13	&	190.1	&	212	&	0.56	\\
5	&	20.7672	&	-38.5622	&	0.0205	&	4.90E+10	&	0.01960	&	1.08	&	2.18	&	1.15E+11	&	7.60E+12	&	65.8	&	159	&	0.14	\\
4	&	25.2701	&	-34.4858	&	0.0128	&	1.14E+10	&	0.01997	&	1.08	&	1.57	&	1.93E+10	&	4.52E+12	&	233.7	&	190	&	0.06	\\
5	&	25.3051	&	-4.2930	&	0.0180	&	3.59E+10	&	0.02266	&	1.09	&	2.05	&	8.05E+10	&	1.92E+13	&	238.4	&	157	&	0.36	\\
5	&	26.2856	&	-13.8575	&	0.0181	&	2.74E+10	&	0.01640	&	1.07	&	2.08	&	6.09E+10	&	1.94E+13	&	318.4	&	175	&	0.29	\\
4	&	28.2577	&	-4.2047	&	0.0164	&	2.91E+10	&	0.02386	&	1.10	&	1.85	&	5.92E+10	&	1.25E+14	&	2110.9	&	331	&	0.52	\\
4	&	28.7924	&	-9.3172	&	0.0173	&	1.91E+10	&	0.02394	&	1.10	&	1.95	&	4.10E+10	&	1.08E+14	&	2635.4	&	265	&	0.7	\\
4	&	29.5184	&	-25.5656	&	0.0144	&	3.66E+10	&	0.01053	&	1.04	&	1.98	&	7.56E+10	&	2.34E+13	&	309.5	&	256	&	0.16	\\
5	&	30.2063	&	-32.1871	&	0.0183	&	4.08E+10	&	0.01977	&	1.08	&	2.33	&	1.03E+11	&	1.17E+14	&	1138.0	&	288	&	0.64	\\
4	&	30.6148	&	-29.3072	&	0.0169	&	3.96E+10	&	0.01656	&	1.07	&	2.15	&	9.10E+10	&	1.34E+14	&	1471.9	&	289	&	0.73	\\
4	&	31.4930	&	-6.8843	&	0.0169	&	1.60E+10	&	0.01961	&	1.08	&	1.83	&	3.16E+10	&	3.71E+13	&	1172.2	&	271	&	0.23	\\
4	&	31.5281	&	-23.1068	&	0.0178	&	5.10E+10	&	0.01674	&	1.07	&	2.12	&	1.16E+11	&	9.35E+13	&	808.6	&	316	&	0.43	\\
4	&	34.8054	&	-38.3532	&	0.0167	&	2.95E+10	&	0.02016	&	1.08	&	1.83	&	5.85E+10	&	4.24E+12	&	72.4	&	55	&	0.64	\\
7	&	36.4329	&	-11.4888	&	0.0153	&	3.37E+10	&	0.02209	&	1.09	&	1.75	&	6.44E+10	&	6.41E+12	&	99.6	&	181	&	0.09	\\
9	&	39.0671	&	-13.1909	&	0.0150	&	5.33E+10	&	0.03001	&	1.13	&	1.68	&	1.01E+11	&	4.99E+13	&	494.9	&	243	&	0.39	\\
9	&	40.8864	&	-25.7116	&	0.0229	&	1.15E+11	&	0.01903	&	1.08	&	3.60	&	4.45E+11	&	8.34E+13	&	187.4	&	219	&	0.79	\\
8	&	41.4914	&	-31.8348	&	0.0221	&	1.13E+11	&	0.02305	&	1.10	&	2.55	&	3.17E+11	&	1.66E+14	&	523.7	&	393	&	0.49	\\
5	&	42.5306	&	-8.9778	&	0.0178	&	2.59E+10	&	0.02907	&	1.12	&	2.04	&	5.93E+10	&	9.15E+13	&	1542.9	&	274	&	0.56	\\
14	&	46.6818	&	-10.4633	&	0.0155	&	8.66E+10	&	0.07630	&	1.36	&	1.67	&	1.96E+11	&	1.17E+14	&	597.3	&	422	&	0.3	\\
5	&	47.4664	&	-25.3934	&	0.0212	&	2.99E+10	&	0.01680	&	1.07	&	2.33	&	7.45E+10	&	3.07E+12	&	41.2	&	113	&	0.11	\\
4	&	48.1498	&	-7.5486	&	0.0177	&	3.10E+10	&	0.06842	&	1.31	&	2.15	&	8.76E+10	&	7.86E+13	&	896.8	&	203	&	0.87	\\
10	&	49.0948	&	-4.7276	&	0.0079	&	2.40E+10	&	0.06228	&	1.28	&	1.26	&	3.87E+10	&	8.08E+13	&	2087.3	&	387	&	0.25	\\
4	&	52.7168	&	-4.8493	&	0.0198	&	3.00E+10	&	0.04634	&	1.20	&	2.24	&	8.10E+10	&	8.13E+13	&	1004.3	&	235	&	0.07	\\
4	&	53.8719	&	-18.4927	&	0.0142	&	2.26E+10	&	0.06259	&	1.28	&	2.06	&	5.97E+10	&	1.94E+12	&	32.5	&	67	&	0.2	\\
11	&	55.2961	&	-4.5489	&	0.0138	&	1.01E+11	&	0.06435	&	1.29	&	1.58	&	2.07E+11	&	3.45E+13	&	166.7	&	216	&	0.34	\\
5	&	149.4540	&	-2.7081	&	0.0210	&	5.70E+10	&	0.04201	&	1.18	&	2.13	&	1.43E+11	&	2.49E+13	&	173.6	&	308	&	0.12	\\
7	&	150.3229	&	-6.2040	&	0.0164	&	4.93E+10	&	0.03796	&	1.16	&	1.78	&	1.02E+11	&	3.36E+13	&	329.0	&	249	&	0.25	\\
5	&	151.6062	&	-2.4202	&	0.0202	&	3.65E+10	&	0.04465	&	1.19	&	2.37	&	1.03E+11	&	2.66E+12	&	25.8	&	58	&	0.36	\\
5	&	155.7194	&	-2.3807	&	0.0186	&	5.69E+10	&	0.04830	&	1.21	&	1.93	&	1.33E+11	&	4.88E+13	&	366.3	&	218	&	0.47	\\
6	&	160.6563	&	-10.2180	&	0.0077	&	1.29E+10	&	0.00910	&	1.04	&	1.38	&	1.77E+10	&	6.25E+13	&	3522.1	&	245	&	0.48	\\
4	&	162.5583	&	-12.2600	&	0.0154	&	1.79E+10	&	0.03559	&	1.15	&	1.86	&	3.84E+10	&	2.77E+13	&	721.7	&	248	&	0.21	\\
4	&	164.2253	&	-9.5503	&	0.0270	&	6.25E+10	&	0.03783	&	1.16	&	3.15	&	2.29E+11	&	2.44E+13	&	106.6	&	191	&	0.31	\\
17	&	165.7276	&	-14.5042	&	0.0144	&	1.09E+11	&	0.00944	&	1.04	&	1.52	&	1.66E+11	&	2.97E+14	&	1787.6	&	475	&	0.6	\\
5	&	168.0586	&	-13.6485	&	0.0171	&	3.23E+10	&	0.07506	&	1.35	&	1.93	&	8.42E+10	&	1.25E+14	&	1484.7	&	351	&	0.46	\\
4	&	170.6570	&	-13.4858	&	0.0173	&	5.77E+10	&	0.03418	&	1.15	&	1.85	&	1.22E+11	&	7.72E+13	&	630.9	&	244	&	0.59	\\
\noalign{\smallskip} 
\hline	    
\normalsize 
\end{tabular} 
\end{flushleft} 
\label{} 
\end{table*}

\begin{table*} 
\caption[]{Same as Table 1.}
\begin{flushleft} 
\begin{tabular}{crrccrccccrrr} 
\hline 
\noalign{\smallskip} 
N	&	$\alpha$ 	&	$\delta$ 	&	z	&
	Raw Lum.	&	E(B-V)	&	ext.	&	comp.	&
	Tot. lum.	&	Mass	&	M/L	&	
$\sigma$	&	Rvir	\\
\hline 
\noalign{\smallskip} 
4	&	171.4233	&	-11.2592	&	0.0174	&	2.83E+10	&	0.04699	&	1.21	&	1.99	&	6.79E+10	&	3.97E+13	&	585.1	&	165	&	0.66	\\
4	&	175.3333	&	-11.9775	&	0.0174	&	4.43E+10	&	0.03384	&	1.14	&	1.97	&	9.98E+10	&	1.35E+14	&	1352.6	&	395	&	0.39	\\
6	&	179.4168	&	-20.3306	&	0.0222	&	4.46E+10	&	0.04524	&	1.20	&	2.31	&	1.23E+11	&	4.77E+13	&	386.6	&	272	&	0.29	\\
4	&	184.8120	&	-12.4777	&	0.0144	&	2.48E+10	&	0.03033	&	1.13	&	1.67	&	4.14E+10	&	1.98E+13	&	477.9	&	127	&	0.56	\\
4	&	190.5120	&	-20.5779	&	0.0224	&	5.86E+10	&	0.05480	&	1.24	&	2.45	&	1.79E+11	&	8.54E+13	&	478.2	&	313	&	0.4	\\
4	&	191.9568	&	-4.7924	&	0.0148	&	4.57E+10	&	0.01745	&	1.07	&	1.93	&	8.82E+10	&	1.28E+14	&	1450.5	&	326	&	0.54	\\
5	&	192.5408	&	-26.9663	&	0.0115	&	2.64E+10	&	0.07663	&	1.36	&	2.16	&	5.69E+10	&	2.71E+13	&	476.1	&	166	&	0.45	\\
26	&	192.5900	&	-13.3802	&	0.0148	&	2.04E+11	&	0.05606	&	1.25	&	1.47	&	3.01E+11	&	3.63E+14	&	1207.8	&	466	&	0.76	\\
4	&	193.6383	&	-20.2686	&	0.0229	&	4.89E+10	&	0.06569	&	1.30	&	2.66	&	1.69E+11	&	5.99E+12	&	35.5	&	223	&	0.06	\\
4	&	195.6448	&	-4.6870	&	0.0103	&	1.72E+10	&	0.01346	&	1.06	&	1.55	&	2.66E+10	&	2.04E+12	&	76.7	&	79	&	0.15	\\
9	&	199.7299	&	-16.4609	&	0.0229	&	9.04E+10	&	0.08047	&	1.38	&	2.41	&	3.00E+11	&	9.28E+13	&	309.1	&	291	&	0.5	\\
5	&	200.9558	&	-11.9613	&	0.0223	&	6.17E+10	&	0.03190	&	1.14	&	2.31	&	1.62E+11	&	3.86E+13	&	238.4	&	238	&	0.31	\\
17	&	201.6564	&	-20.3991	&	0.0183	&	1.10E+11	&	0.08540	&	1.41	&	1.88	&	2.06E+11	&	5.93E+13	&	287.5	&	242	&	0.46	\\
4	&	201.9665	&	-1.8782	&	0.0136	&	2.51E+10	&	0.03450	&	1.15	&	1.69	&	4.25E+10	&	4.04E+13	&	951.6	&	262	&	0.27	\\
4	&	201.9803	&	-21.4628	&	0.0243	&	2.69E+10	&	0.09882	&	1.48	&	2.52	&	1.00E+11	&	2.59E+12	&	25.8	&	93	&	0.14	\\
11	&	202.8497	&	-24.7016	&	0.0159	&	7.93E+10	&	0.06650	&	1.30	&	1.74	&	1.38E+11	&	1.39E+14	&	1007.2	&	278	&	0.82	\\
4	&	229.1948	&	-13.2268	&	0.0073	&	1.25E+10	&	0.03372	&	1.14	&	1.43	&	1.78E+10	&	1.56E+12	&	87.6	&	104	&	0.07	\\
4	&	318.7207	&	-23.0257	&	0.0269	&	6.63E+10	&	0.05581	&	1.25	&	3.3	&	2.73E+11	&	1.06E+13	&	38.8	&	228	&	0.09	\\
32	&	329.3656	&	-33.4588	&	0.0087	&	1.01E+11	&	0.03357	&	1.14	&	1.26	&	1.46E+11	&	2.54E+13	&	174.4	&	166	&	0.42	\\
6	&	331.6321	&	-27.8423	&	0.0238	&	1.49E+11	&	0.02030	&	1.08	&	2.64	&	4.28E+11	&	8.80E+12	&	20.6	&	209	&	0.09	\\
5	&	332.2561	&	-27.0691	&	0.0086	&	6.75E+09	&	0.02354	&	1.10	&	1.32	&	9.78E+09	&	5.11E+11	&	52.3	&	30	&	0.26	\\
4	&	332.4736	&	-22.8598	&	0.0178	&	4.14E+10	&	0.02640	&	1.11	&	2.5	&	1.15E+11	&	2.63E+12	&	22.9	&	52	&	0.44	\\
4	&	332.6806	&	-30.0202	&	0.0139	&	4.61E+10	&	0.01484	&	1.06	&	2.39	&	1.17E+11	&	6.94E+13	&	593.3	&	262	&	0.46	\\
4	&	332.9872	&	-27.7711	&	0.0178	&	5.87E+10	&	0.01950	&	1.08	&	2.74	&	1.74E+11	&	1.01E+13	&	58.1	&	122	&	0.31	\\
7	&	333.7648	&	-21.2985	&	0.0087	&	1.61E+10	&	0.02895	&	1.12	&	1.33	&	2.40E+10	&	5.09E+11	&	21.2	&	59	&	0.07	\\
4	&	335.6633	&	-31.4416	&	0.0285	&	6.64E+10	&	0.01282	&	1.05	&	3.9	&	2.72E+11	&	6.78E+13	&	248.9	&	295	&	0.36	\\
16	&	341.0783	&	-22.2565	&	0.0107	&	9.25E+10	&	0.02412	&	1.10	&	1.38	&	1.41E+11	&	2.88E+13	&	204.9	&	130	&	0.78	\\
4	&	343.0948	&	-34.2753	&	0.0291	&	6.90E+10	&	0.01265	&	1.05	&	3.86	&	2.80E+11	&	1.57E+12	&	5.6	&	93	&	0.08	\\
14	&	356.4653	&	-28.3663	&	0.0282	&	2.60E+11	&	0.01613	&	1.07	&	3.79	&	1.05E+12	&	3.56E+13	&	33.9	&	315	&	0.16	\\
4	&	357.4948	&	-29.5501	&	0.0291	&	5.24E+10	&	0.01603	&	1.07	&	3.97	&	2.22E+11	&	1.62E+13	&	73.1	&	143	&	0.36	\\
\noalign{\smallskip} 
\hline	    
\normalsize 
\end{tabular} 
\end{flushleft} 
\label{} 
\end{table*}

\begin{table} 
\caption[]{Various relations between M/L ratio and velocity dispersions for 
our sample. The first column describes the selection (number of galaxies per 
group or cluster sample of Adami et al. 1998), the second column is the slope 
of the linear relation and the third column is the constant term of the linear 
relation. The fourth column is the dispersion (in percents) of the relations
for groups.}
\begin{flushleft} 
\begin{tabular}{rrrr} 
\hline 
\noalign{\smallskip} 
Selection	&	slope 	&	constant & dispersion\\
\hline 
\hline
\noalign{\smallskip} 
All SSRS2 groups & 3.79$\pm$0.64 & 294$\pm$570 & 8$\%$	\\
\hline 
SSRS2 groups with  & 3.57$\pm$0.89 & 462$\pm$431 & 6$\%$	\\
more than 7 galaxies &  & & 	\\
\hline 
SSRS2 groups with & 3.90$\pm$1.17 & 579$\pm$470 & 4$\%$	\\
more than 8 galaxies & & & \\
\hline 
Cluster sample & 0.58$\pm$0.17 & 12$\pm$236 & -	\\
\noalign{\smallskip} 
\hline	    
\normalsize 
\end{tabular} 
\end{flushleft} 
\label{} 
\end{table}

\section{Conclusion}

We demonstrated that we are able to detect similar populations of groups
with our method compared to classical friend-of-friend algorithms. The
groups detected with our method have the same physical properties than those
detected by Merchan et al. (2000): nearly same mass distribution (most of
our groups are in the range [$5.10^{12}$;$4.10^{14}$] solar masses),
same abundances up to cz$\sim $8000 km/s (and, therefore, same completeness
limit: $\sim $5000 km/s and abundances in agreement with Press $\&$ Schechter
models), and same 2-point correlation function (modeled by a power law: $%
(r/r_0)^\gamma $ with $r_0=7h^{-1}Mpc$ and $\gamma =-1.79$).

This redshift completeness limit close to 5000 km/s is directly related to
the magnitude limit of the catalog: m$_{SSRS2}\sim B\sim 15.5$.
Extrapolating these limits assuming the typical depth of the new generation
of redshift surveys using 10 meter class telescopes (Keck and VLT) leads to
a redshift completeness limit of z$\sim $0.7 for the groups in these
samples.

We found for the SSRS2 catalog a similar luminosity function for nearby group 
galaxies to that of
Zabludoff et al. (2000): M$^{*}$=(-19.99$\pm $0.36)+5logh and $\alpha $=-1.46%
$\pm $0.17. We computed a similar mass to light ratio compared to Carlberg
et al. (2001) (median value of 250 hM$_{\odot }$/L$_{\odot }$ in the B
SSRS2 magnitude passband) and we deduced a similar relation between the mass
to light ratio and velocity dispersion ($M/L=(3.79\pm 0.64)\sigma
-(294\pm 570)$). This relation is qualitatively similar to that detected for
rich clusters of galaxies (Adami et al. 1998), but with a significantly
steeper slope. The more massive the group (or the cluster), the higher the
mass to light ratio, and therefore, the larger the amount of dark matter
inside group (and cluster). Another explanation is a significant stripping
of the gas of the galaxies in rich groups as opposed to poorer groups.
Finally, we detected a fundamental plane for these groups ($L\propto
R_{Vir}^{2.26\pm 1.39}\times \sigma ^{2.93\pm 1.64}$ for groups with more
than 8 galaxies) but much less narrow and barely significant compared to
clusters of galaxies. We conclude that a mean relation between L, R$_{Vir}$
and $\sigma $ probably exists for groups, but this relation is much less
well defined compared to clusters.

\begin{acknowledgements}

{The authors thank the referee for useful comments, all the VIRMOS team
in France and Italy for help and F. Durret for a careful reading of the
manuscript.}

\end{acknowledgements}

\end{document}